\documentclass[12pt]{iopart}
\usepackage{longtable,booktabs}
\usepackage{graphicx}
\usepackage{dcolumn}
\usepackage{bm}
\usepackage{color}
\usepackage{upgreek}
\usepackage{threeparttable}
\usepackage[numbers,sort&compress]{natbib}
\usepackage{multirow}
\usepackage{epstopdf}
\usepackage{array}
\usepackage{bm}
\usepackage {indentfirst}
\usepackage{caption}
\expandafter\let\csname equation*\endcsname\relax
\expandafter\let\csname endequation*\endcsname\relax
\usepackage{amsmath}

\makeatletter

\newcommand{\Rmnum}[1]{\expandafter\@slowromancap\romannumeral #1@}
\makeatother
\usepackage{setspace}

\usepackage{color}

\begin{document}

\title[]{Be optical lattice clocks with the fractional Stark shift up to the level of 10$^{-19}$}

\author{Lei Wu, Xia Wang, Ting Wang, Jun Jiang$^{\dagger}$, Chenzhong Dong}
\address{ Key Laboratory of Atomic and Molecular Physics and Functional Materials of Gansu Province, College of Physics and Electronic Engineering, Northwest Normal University, Lanzhou 730070, China}
\ead{phyjiang@yeah.net}

\begin{abstract}
The energy levels and electric dipole ($E1$) matrix elements of the ground state and low-lying excited states of Be atoms are calculated using the relativistic configuration interaction plus core polarization (RCICP) method. The static and dynamic $E1$, magnetic dipole ($M1$) and electric quadrupole ($E2$) polarizabilities as well as the hyperpolarizabilities of the $2s^2~^1S_0$ and $2s2p~^3P_0$ states are determined. Two magic wavelengths, 300.03 and 252.28 nm, of $2s^2~^1S_0 \rightarrow 2s2p~^3P_0$ clock transition are found. Then, the multipolar and nonlinear Stark shifts of the clock transition at the magic wavelength are discussed in detail. We find that when the laser intensity $I$ is in the range of 14.3 $\sim$ 15.9 kW/cm$^{2}$ and the detuning $\delta$ (the frequency detuning of the lattice laser frequency relative to the magic frequency) is in the range of 40.7 $\sim$ 40.9 MHz, the fractional Stark shifts of the clock transition are less than 1.0 $\times$ 10$^{-18}$. While, when $I$ is in the range of 15.01 $\sim$ 15.46 kW/cm$^{2}$ and $\delta$ is in the range of 40.73 $\sim$ 40.76 MHz, the fractional Stark shifts are lower than 1.0 $\times$ 10$^{-19}$. 
\end{abstract}
\newcommand\keywords[1]{\textbf{Keywords}: polarizability, magic wavelengths, multipolar and nonlinear Stark shifts}
%\maketitle
\section{Introduction}

In the past few decades, with the rapid development of laser cooling and trapping techniques, extraordinary advancements in optical atomic clock accuracy and stability have been demonstrated~\cite{Ludlow2015,Cryogenic2015,Bothwell2022,Brewer2019,Nicholson2015,McGrew2018}. The high-accuracy optical clocks can be used for performing precision measurements of fundamental physical constants~\cite{Yamanaka2015,Bregolin2017}, testing the local Lorentz invariance~\cite{Shaniv2018,Pihan2017}, exploring variations of fine structure constant $\alpha$ with time~\cite{Godun2014,Safronova2018}, probing dark matter and dark energy~\cite{Arvanitaki2015,Roberts2017}, detecting gravitational waves~\cite{Kolkowitz2016}, and detecting new forces beyond the standard model of particle physics~\cite{Kolkowitz2016,Roberts2017}.

Meticulously studying the interaction between laser fields and atoms has become the core of developing ultra-higher-precision optical clocks. The interaction between laser fields and atoms can cause ac Stark shift, which would affect the accuracy of the measurement of relevant atomic parameters. To reduce the impact of the ac Stark shift, magic-wavelength trapping was introduced in Refs.~\cite{Katori1999,Ye1999,Nicholson2015,McGrew2018}. The dynamic electric dipole ($E1$) polarizabilities of a given pair of energy levels are the same as each other in the magic-wavelength trapping, and the second-order Stark shift is eliminated. However, when the accuracy of the optical lattice clock is at the level of $10^{-18}$, the multipolar (electric quadrupole $E2$ and magnetic dipole $M1$) and nonlinear Stark shifts are non-negligible~\cite{Taichenachev2008,Ovsiannikov2013,Katori2015,Ovsiannikov2016,Wu2020}. These Stark shifts are related to the $E2$ and $M1$ polarizabilities as well as hyperpolarizability~\cite{Taichenachev2008,Katori2009, Katori2015,Ovsiannikov2013,Ovsiannikov2016,Porse2018,Ushijima2018,Wu2019}. For pursuing much higher precision, the effects on the systematic uncertainty of optical clocks from the multipolar and nonlinear Stark shifts need to be evaluated~\cite{Taichenachev2008,Ovsiannikov2013,Katori2015,Ovsiannikov2016,Porse2018,Ushijima2018}.

Be atoms have been proposed as one of the potential candidates for developing ultra-high-precision optical clocks due to their unique properties~\cite{Mitroy2010}. Compared to other neutral atoms, the blackbody radiation shifts at room temperature of $2s2p~^1S_0$ $\rightarrow$ $2s^2~^3P_0$ clock transition is about 1.7$\times$10$^{-17}$~\cite{NIST_ASD}, which is one or two orders of magnitude smaller than that of the Mg, Sr and Yb atoms ~\cite{Kulosa2015,Middelmann2012,Sherman2012}. The transition wavelength, 454.997 nm, lies in the optical range and the natural line width is about 0.068 Hz~\cite{NIST_ASD}. Consequently, the quality factor, the ratio of the transition frequency to the natural line width, of this line is about 1$\times$10$^{16}$~\cite{NIST_ASD}. Magic wavelengths for simultaneous trapping of the ground and metastable states were calculated by Mitroy~\cite{Mitroy2010}. As far as we know, there are no reported values for the dynamic $M1$ and $E2$ polarizabilities as well as hyperpolarizabilities for Be atoms. Therefore, there is also no theoretical analysis of the influence of $E2$ and $M1$ interactions and nonlinear Stark shifts on the accuracy of optical lattice clocks.

In this paper, the energy levels and $E1$ matrix elements of the ground state and low-lying excited states for Be atoms are calculated using the relativistic configuration interaction plus core polarization (RCICP) method. The static and dynamic $E1$, $M1$ and $E2$ polarizabilities as well as the hyperpolarizabilities of the $2s^2~^1S_0$ and $2s2p~^3P_0$ states are further determined. Then, the laser intensities and detunings of the optical lattice to improve the fractional Stark shifts of the clock transition to 10$^{-19}$-10$^{-20}$ are analyzed in detail. Atomic units ($m_{e}$ = 1, $e$ = 1, $\hbar$ = 1) are used throughout the paper unless stated otherwise. The speed of light is taken to be 137.035 999 1 in our calculations.

\section{Theoretical method}

The key strategy of the RCICP method is to partition a Be atom into a Be$^{2+}$ core plus two valence electrons. The calculation is separated into three steps. The first step involves a Dirac-Fock (DF) calculation on the core of Be$^{2+}$ ions. The second step is to obtain the single-electron wave function of valence orbitals. The single-particle orbitals are written as a linear combination of the S-spinors~\cite{Grant1988, Grant1989, Grant94, Grant2007} which can be regarded as a relativistic generalization of the Slater-type orbitals. The third step is to diagonalize the Hamiltonian matrix in two-valence-electrons configuration space. The effective Hamiltonian of the two valence electrons is written as
\begin{equation}
	\label{tp}
	H=\sum\limits_{i=1}^{2}(c \bm{\alpha} \cdot 
	\bm{p}+(\beta-1) c^{2}+V_{\rm{core}}(\bm{r_{i}}))
	+\frac{1}{r_{12}}+V_{\rm{p2}}(\bm{r_{1}},\bm{r_{2}}),
\end{equation}
where the summation part represents single-electron Hamiltonian,
$\bm{\alpha} $ and $ \beta $ are the $ 4\times 4 $ Dirac matrices,
$\bm{p}$ is the momentum operator, $ c $ is the speed of light,
and $\bm{r}$ is the position vector of the valence electron.
Moreover, $V_{\rm{core}}(\bm{r})$ is given by
\begin{equation}
V_{\rm{core}}(\bm{r}) = -\frac{Z}{r} + V_{\rm{dir}}(\bm{r}) +
V_{\rm{exc}}(\bm{r}) + V_{\rm{p1}}(\bm{r}).
\end{equation}
Here, $ Z $ is atomic number, $ r $ is the distance of the
valence electron with respect to the origin of the coordinates.
$V_{\rm{dir}}(\bm{r})$ and $V_{\rm{exc}}(\bm{r}) $ denote the
direct and exchange interactions of the valence electron with the 
core electrons, respectively. The $\ell$- and $j$-dependent
one-electron polarization potential $ V_{\rm{p1}}(\bm {r}) $ can be written as~\cite{Mitroy2010a}
\begin{equation}
\label{p1}
V_{\rm{p1}}(\bm{r}) = -\sum_{k=1}^{3}\frac{\alpha_{\rm{core}}^{k}}{2r^{2(k+1)}}
\sum_{\ell,j}g_{\ell,j}^{2}(r)|\ell j\rangle\langle \ell j|.
\end{equation}
The two-electron polarization potential is written as~\cite{Mitroy2010a}
\begin{equation}
\label{p2}
V_{p2}(\bm{r}_1,\bm{r}_2) = -\sum_{k=1}^{3} \frac
{\alpha_{\mathrm{core}}^{(k)}}{r_1^{2(k+1)}r_2^{2(k+1)}}
\sum_{\ell,j} g_{k,\ell,j}(r_1)g_{k,\ell,j}(r_2)(\bm{r}_1
\cdot \bm{r}_2),
\end{equation}
where $ \ell $ and $j$ are the orbital and total angular momenta, respectively. 
$\alpha_{\rm{core}}^{k}$ is the $k$th-order static polarizabilities of the core electrons 
($\alpha_{\mathrm{core}}^{(1)}$ = 5.227$\times10^{-2}$ a.u.,
$\alpha_{\mathrm{core}}^{(2)}$ = 1.532$\times10^{-2}$ a.u. and
$\alpha_{\mathrm{core}}^{(3)}$ = 1.135$\times10^{-2}$ a.u.~\cite{Bhatia1997}
for Be$^{2+}$ ions), and $g_{k,\ell,j}^{2}(r) = 1-\exp(-r^{2(k + 2)}/\rho_{\ell,j}^{2(k + 2)})$.
The cutoff parameters $ \rho_{\ell,j} $ are tuned to reproduce 
the binding energies of the ground state and some low-lying excited
states, which are listed in Table~\ref{rho}. The effective Hamiltonian
of the valence electron is diagonalized within a large S-spinor and
L-spinor basis~\cite{Grant2007,Grant2000}. L-spinors can be regarded
as a relativistic generalization of the Laguerre-type orbitals. 

The energy levels of some low-lying states of Be atoms are
listed in Table~\ref{level}, and compared with the National
Institute of Science and Technology (NIST) tabulation~\cite{NIST_ASD}.
It can be found that the present RCICP results are in good
agreement with those from the NIST tabulation, and the difference is no more than 0.07\%.

\begin{table}[ht]
	\centering
	\footnotesize\rm
	\captionsetup{font=footnotesize,labelfont=bf}
	\caption{\label{rho} The radial cutoff parameter $\rho_{\ell,j}$ of the polarization potential of Be atoms.}
	\setlength{\tabcolsep}{12mm}{
		\begin{tabular}{ccc} 
			\\\hline  \hline                                     
			States&                $j$&          $\rho_{\ell,j} $ (a. u.)           \\ \hline
			$s$   &                1/2&              0.9587                 \\
			$p$   &                1/2&              0.8695                 \\
			      &                3/2&              0.8672                 \\
			$d$   &                3/2&              1.3305                 \\
			      &                5/2&              1.3286                 \\\hline \hline           
		\end{tabular} }                                              
\end{table}
\begin{table}[ht]
	\centering
	\footnotesize\rm
	\captionsetup{font=footnotesize,labelfont=bf}
	\caption{\label{level} Energy levels (cm$^{-1}$) for some of the
		low-lying states of the Be atoms, which are given
		relatively to the core Be$^{2+}$. The relative differences between
		the present RCICP results and the National Institute of Science and
		Technology (NIST) tabulation~\cite{NIST_ASD} are listed as Diff.}
	\setlength{\tabcolsep}{11.6mm}{
		\begin{tabular}{cccc}
			    \\\hline  \hline   
			State&                             RCICP&            NIST~\cite{NIST_ASD}&        Diff.               \\ \hline
			$2s^2$  $^1S_{0}$&             $-$222065.82&          $-$222075.47&           $-$0.004\%              \\
			$2s3s$  $^1S_{0}$&             $-$167384.93&          $-$167398.13&           $-$0.008\%              \\
			$2p^2$  $^3P_{0}$&             $-$162326.52&          $-$162381.88&           $-$0.034\%              \\
			$2s4s$  $^1S_{0}$&             $-$156783.08&          $-$156830.05&           $-$0.030\%              \\
			$2s5s$  $^1S_{0}$&             $-$152676.20&          $-$152753.18&           $-$0.050\%              \\
			$2s6s$  $^1S_{0}$&             $-$150653.17&          $-$150754.23&           $-$0.067\%              \\
			$2s2p$  $^3P_{0}$&             $-$200086.53&          $-$200097.17&           $-$0.005\%              \\
			$2s3p$  $^3P_{0}$&             $-$163129.57&          $-$163168.01&           $-$0.024\%              \\
			$2s4p$  $^3P_{0}$&             $-$155212.55&          $-$155263.60&           $-$0.033\%              \\
			$2s5p$  $^3P_{0}$&             $-$151967.42&          $-$152010.08&           $-$0.028\%              \\
			$2s6p$  $^3P_{0}$&             $-$150304.45&          $-$150344.84&           $-$0.027\%              \\
			$2s2p$  $^3P_{1}$&             $-$200089.21&          $-$200096.55&           $-$0.004\%              \\
			$2s2p$  $^1P_{1}$&             $-$179482.17&          $-$179510.03&           $-$0.016\%              \\
			$2s3p$  $^3P_{1}$&             $-$163109.39&          $-$163168.01&           $-$0.036\%              \\
			$2s3p$  $^1P_{1}$&             $-$161806.87&          $-$161888.04&           $-$0.050\%              \\
			$2s4p$  $^3P_{1}$&             $-$155183.49&          $-$155263.61&           $-$0.052\%              \\
			$2s4p$  $^1P_{1}$&             $-$154952.51&          $-$155040.69&           $-$0.057\%              \\
			$2s5p$  $^3P_{1}$&             $-$151947.39&          $-$152010.08&           $-$0.041\%              \\
			$2s5p$  $^1P_{1}$&             $-$151878.95&          $-$151954.90&           $-$0.050\%              \\
			$2s2p$  $^3P_{2}$&	           $-$200089.78&	      $-$200094.21&	          $-$0.002\%              \\
			$2s3p$  $^3P_{2}$&	           $-$163118.88&	      $-$163167.64&	          $-$0.030\%              \\
			$2s4p$  $^3P_{2}$&	           $-$155199.36&	      $-$155263.60&	          $-$0.041\%              \\
			$2s4f$  $^3F_{2}$&	           $-$153781.72&	      $-$153834.44&	          $-$0.034\% 			  \\
			$2s3s$  $^3S_{1}$&	           $-$169987.75&	      $-$169994.54&	          $-$0.004\%              \\
			$2p^2$  $^3P_{1}$&	           $-$162364.74&	      $-$162380.49&	          $-$0.010\%              \\
			$2s3d$  $^3D_{1}$&	           $-$159976.31&	      $-$160021.74&	          $-$0.028\%              \\
			$2s4s$  $^3S_{1}$&	           $-$157510.31&	      $-$157569.00&	          $-$0.037\%              \\
			$2s4d$  $^3D_{1}$&	           $-$154082.45&	      $-$154133.85&	          $-$0.033\%              \\
			$2s5s$  $^3S_{1}$&	           $-$152992.49&	      $-$153065.31&	          $-$0.048\%              \\
			$2s5d$  $^3D_{1}$&	           $-$151418.88&	      $-$151471.72&	          $-$0.035\% 			  \\
			$2p^2$  $^1D_{2}$&	           $-$165187.86&	      $-$165192.93&	          $-$0.003\%              \\
			$2p^2$  $^3P_{2}$&	           $-$162369.64&	      $-$162378.50&	          $-$0.005\%              \\
			$2s3d$  $^3D_{2}$&	           $-$159994.17&	      $-$160021.74&	          $-$0.017\%              \\
			$2s3d$  $^1D_{2}$&	           $-$157576.19&	      $-$157647.08&	          $-$0.045\%              \\
			$2s4d$  $^3D_{2}$&	           $-$154056.36&	      $-$154133.83&	          $-$0.050\%              \\
			$2s4d$  $^1D_{2}$&	           $-$153194.25&	      $-$153294.53&	          $-$0.065\%              \\\hline  \hline                  
		\end{tabular}  }                                                     
\end{table}

\section{Results and discussion}
\subsection{$E1$, $E2$ and $M1$ transitions matrix elements}

The $E1$ transition matrix elements are calculated with a
modified dipole transition operator given by~\cite{Mitroy88,Marinescu94,Caves72,Hameed68,Hafner78}
\begin{equation}
\label{E1}
\bm{D}=\bm{r}-[1-\mathrm{exp}(-\frac{r^{6}}{\rho^{6}})]^{1/2}\frac{\alpha_{\rm{core}}^{1} \bm{r}}{r^{3}}.
\end{equation}
The cutoff parameter $\rho$ is 1.0522 a.u., generated as $\rho=\frac{1}{6}(2\rho_{s_{1/2}}+\rho_{p_{1/2}}+\rho_{p_{3/2}}+\rho_{d_{3/2}}+\rho_{d_{5/2}})$.

Table~\ref{matrix} lists the presently calculated $E1$ reduced matrix elements for transitions between some low-lying states, along with a comparison with some available theoretical~\cite{Mitroy2010c,Chen1998,Charlotte2004,begue1998,Fuhr2010,wang2018} and experimental results~\cite{Martinson1974,Irving1999,Kerkhoff1980,Bromander1971}. For the $2s^2~^1S_0 \rightarrow 2s2p~^1P_1$ resonant transition, which is dominant contributing to the polarizability of the ground state, the present RCICP result is in good agreement with experimental~\cite{Martinson1974,Irving1999} and theoretical~\cite{Mitroy2010c,Chen1998,Charlotte2004,begue1998,Fuhr2010,wang2018} results. The differences are less than 1.5\%. For $2s2p~^3P_0 \rightarrow 2p^2~^3P_1$ and $2s2p~^3P_0 \rightarrow 2s3d~^3D_1$ transitions, the reduced matrix elements are larger than 1.0 a.u., and the present RCICP results agree very well with the experimental results~\cite{Kerkhoff1980,Bromander1971}. The agreement is better than 1\%. Moreover, there are no comparable experimental values for the transitions of reduced matrix elements of less than 1.0 a.u., except for the $2s2p~^3P_0 \rightarrow 2s3s~^3S_1$ transition. The present results agree well with other theoretical results~\cite{Mitroy2010c,Chen1998,Charlotte2004,begue1998,Fuhr2010,wang2018}. For $2s2p~^3P_0 \rightarrow 2s3s~^3S_1$ transition, the difference between the present result and experimental results~\cite{Kerkhoff1980,Bromander1971} is no more than 4\%.

\begin{table}[ht]
	\centering
	\footnotesize\rm
	\captionsetup{font=footnotesize,labelfont=bf}
	\caption{\label{matrix} Reduced $E1$ matrix elements (a.u.) for some of principal transitions of Be atoms.}
	\setlength{\tabcolsep}{0mm}{	
		\begin{tabular}{ccccccccc}
			\\\hline  \hline
			Transition&          This work&  CICP~\cite{Mitroy2010c}&  BCICP~\cite{Chen1998}&    BCIBP~\cite{Charlotte2004}&  TDGI~\cite{begue1998}&    MCHF~\cite{Fuhr2010}&   HFR~\cite{wang2018}&              Exp.      \\ \hline
			$2s^2~^1S_0 \rightarrow 2s2p~^1P_1$&    3.2606&      3.2597&      3.260&      3.262&    3.270&       3.256&      3.306&                         3.22(6)~\cite{Martinson1974}   \\
			&            &            &           &           &         &             &          &                         3.29(5)~\cite{Irving1999}     \\
			$2s^2~^1S_0 \rightarrow 2s3p~^1P_1$&    0.2212&        0.2179&      0.222&      0.222&         &       0.221&           &                          \\ 
			$2s^2~^1S_0 \rightarrow 2s4p~^1P_1$&    0.0336&            &        0.034&           &         &       0.024&           &                         \\
			$2s^2~^1S_0 \rightarrow 2s5p~^1P_1$&    0.0620&            &        0.062&           &         &       0.057&           &                          \\ 
			$2s2p~^3P_0 \rightarrow 2s3s~^3S_1$&    0.9505&      0.9091&        0.948&      0.961&    0.534&       0.961&      0.954&            0.97(1)~\cite{Kerkhoff1980}    \\
		                                    	&          &            &              &           &         &            &          &            0.99(2)~\cite{Bromander1971}     \\
			$2s2p~^3P_0 \rightarrow 2p^2~^3P_1$&    1.9804&      1.9740&              &      1.972&         &       1.972&     2.045&             1.96(1)~\cite{Bromander1971}                  \\
			$2s2p~^3P_0 \rightarrow 2s3d~^3D_1$&    1.5576&      1.5551&         1.556&      1.568&    1.118&       1.568&     1.503&             1.54(2)~\cite{Kerkhoff1980}      \\
			&          &            &              &           &         &            &          &                       1.54(2)~\cite{Bromander1971}   \\
			$2s2p~^3P_0 \rightarrow 2s4s~^3S_1$&    0.2995&            &         0.300&           &         &       0.300&     0.288&                           \\
			$2s2p~^3P_0 \rightarrow 2s4d~^3D_1$&    0.8276&            &         0.827&           &         &       0.829&     0.737&         \\\hline  \hline                                             
		\end{tabular} }                                           
\end{table}

\begin{table}[ht]
	\centering
	\footnotesize\rm
	\captionsetup{font=footnotesize,labelfont=bf}
	\caption{\label{matrixE2M1} The $M1$ and $E2$ matrix elements (a.u.) for principal transitions of Be atoms.
		The notation a[b] means a$\times10^{b}$.}
	\setlength{\tabcolsep}{1mm}{
	\begin{tabular}{ccclcccc}
		\hline  \hline
		Type&			Transition&	                               This work&       Other studies&               Type&           Transition&	                  This work&       Other studies\\ \hline
		M1&    $2s^2~^1S_0 \rightarrow 2s3s~^3S_1$&                 1.24[$-$5]&         1.29[$-$5]~\cite{Ray1988}& M1&	$2s2p~^3P_0 \rightarrow 2s2p~^3P_1$&             1.43&         1.41~\cite{NIST_ASD} \\
		M1&    $2s^2~^1S_0 \rightarrow 2s4s~^3S_1$&                 6.77[$-$6]&         6.70[$-$6]~\cite{Ray1988}& M1&  	$2s2p~^3P_0 \rightarrow 2s2p~^1P_1$&             1.16[$-$4]&         1.14[$-$4]~\cite{NIST_ASD}\\ 
		M1&    $2s^2~^1S_0 \rightarrow 2s5s~^3S_1$&                 4.51[$-$6]&         4.45[$-$6]~\cite{Ray1988}& M1&  	$2s2p~^3P_0 \rightarrow 2s3p~^3P_1$&             5.24[$-$5]&        \\ 
		M1&    $2s^2~^1S_0 \rightarrow 2s6s~^3S_1$&                 3.52[$-$6]&         3.33[$-$6]~\cite{Ray1988}& M1&  	$2s2p~^3P_0 \rightarrow 2s3p~^1P_1$&             9.71[$-$5]&        \\
		E2&    $2s^2~^1S_0 \rightarrow 2p^2~^1D_2$&                    6.17&                                 & E2&   $2s2p~^3P_0 \rightarrow 2s2p~^3P_2$&             7.25&         7.20~\cite{NIST_ASD} \\
		E2&    $2s^2~^1S_0 \rightarrow 2p^2~^3P_2$&                2.02[$-$1]&                      &            E2&   $2s2p~^3P_0 \rightarrow 2s3p~^3P_2$&             5.43&         \\
		E2&    $2s^2~^1S_0 \rightarrow 2s3d~^3D_2$&                    3.15&                      &            E2&   $2s2p~^3P_0 \rightarrow 2s4p~^3P_2$&             2.04&        \\
		E2&    $2s^2~^1S_0 \rightarrow 2s3d~^1D_2$&                    12.48&         12.50~\cite{Yi2001}  &   E2&   $2s2p~^3P_0 \rightarrow 2s4f~^3F_2$&             3.05&         \\			 
		\\\hline  \hline
	\end{tabular}  }                                           
	\end{table}

Table~\ref{matrixE2M1} lists the presently calculated $E2$ and $M1$ matrix elements for some important transitions and some available theoretical results~\cite{Yi2001,Ray1988,NIST_ASD}. The $M1$ matrix elements are very small, about 10$^{-4}$ $\sim$ 10$^{-6}$, except for the $2s2p~^3P_0 \rightarrow 2s2p~^3P_1$ transition. The present RCICP results agree very well with the other theoretical results~\cite{Yi2001,NIST_ASD}. The differences are less than 2\%. For the $2s2p~^3P_0 \rightarrow 2s2p~^3P_1$ transition, the $M1$ matrix element are four to six orders of magnitude larger than the $M1$ matrix elements of the other $M1$ transitions. For the $E2$ matrix elements, we only found two available theoretical data~\cite{Yi2001,NIST_ASD}, that is for the $2s^2~^1S_0 \rightarrow 2s3d~^1D_2$ and $2s2p~^3P_0 \rightarrow 2s2p~^3P_2$ transitions, to compare with the present results. The present RCICP results agree very well with these two results, and the difference is no more than 1\%.
	
\subsection{Static and dynamic $E1$ polarizabilities}

The dynamic $E1$ polarizability of the state $i$ with the total angular momenta $j=0$ can be given by
\begin{equation}
\label{s}
\alpha_{i}(\omega) = \sum_{n} \frac{f_{i \rightarrow n}}
{\Delta E_{n \rightarrow i}^2-\omega^2},
\end{equation}
where $\Delta E_{n \rightarrow i}$ is the transition energy
and $\omega$ is the laser frequency. When $\omega$ = 0,
Eq. (\ref{s}) is reduced to the static polarizabilities.
The $E1$ oscillator strength $f_{i \rightarrow n}$ is defined as
\begin{equation}
f^{(1)}_{i \to n}=\frac{2|\langle \beta_{n}j_{n}\|\bm{D}\|
	\beta_{i}j_{i}\rangle |^{2}\Delta E_{n \to i}}{3(2j_{i}+1)},
\end{equation}
where $j$ is the total angular momenta and $\beta$ represents
all additional angular momenta in addition to the total angular momenta $j$.

\begin{table}[ht]
	\centering
	\footnotesize\rm
	\captionsetup{font=footnotesize,labelfont=bf}
	\caption{\label{E1 polarizabilities} Static $E1$ polarizabilities $\alpha$ (a.u.) of the $2s^2~^1S_0$ and $2s2p~^3P_0$ states and the breakdown of the contributions of individual transitions for Be atoms.}
	\setlength{\tabcolsep}{7mm}{
		\begin{tabular}{ccccc}
			\\\hline  \hline
			\multicolumn{2}{c}{	$2s^2~^1S_0$}&  &\multicolumn{2}{c}{$2s2p~^3P_0$}   \\ \cline{1-2}  \cline{4-5}	
			Contributions&            $\alpha$&  &        Contributions&           $\alpha$      \\ \hline
			$2s2p~^1P_1$&             36.5299&    &       $2s3s~^3S_1$&            4.3919        \\
			$2s3p~^1P_1$&              0.1189&    &       $2p^2~^3P_1$&            15.2131       \\ 
			$2s4p~^1P_1$&              0.0025&    &       $2s3d~^3D_1$&            8.8504        \\
			$2s5p~^1P_1$&              0.0080&    &       $2s4s~^3S_1$&            0.3082        \\
			&                    &                &        $2s4d~^3D_1$&            2.1782         \\
			Remainds&              1.0757&          &             &            8.1180         \\  
			Core&                        0.0523&     &                  &            0.0523        \\
			Total&                       37.7873&      &                 &            39.1121        \\
			MCDHF~\cite{dong2021}&              37.614&     &         &            39.249        \\
			MCHF~\cite{Themelis1995}&           37.62&      &        &            39.33          \\
			CICP~\cite{Mitroy2010}&             37.73&      &        &            39.04          \\
			TDGI~\cite{Pouchan1998,Claude1998}& 37.62&      &       &            36.08          \\
			Sum-over-states~\cite{Shukla2020}&  36.6&       &        &                           \\  
			Model potential~\cite{Patil2000} &  37.9&      &       &                            \\
			CI+MBPT~\cite{porsev2006}&            37.76&   &            &                             \\
			Hylleraas Weinhold~\cite{Komasa2002}&  37.755&   &       &                            \\
			ECG~\cite{Komasa2001}&                  37.755&    &     &                             \\
			CI~\cite{Bendazzoli2004}&                37.8066&  &     &                             \\
			RCC~\cite{Sahoo2008}&              37.80(47)&      &     &                              \\
			RCC+MBPT~\cite{Singh2013}&         37.86(17)&      &     &                               \\
			Semi-empirical~\cite{Mitroy2003}&         37.69&    &    &                              \\
			Ab initio~\cite{BANERJEE2010}&         38.12&        &   &                             \\
			VP+CI~\cite{Figari1983}&                37.59&      &    &                             \\\hline  \hline
		\end{tabular} }                                               
\end{table}

Table~\ref{E1 polarizabilities} lists the presently calculated static $E1$ polarizabilities of the $2s^2~^1S_0$ and $2s2p~^3P_0$ states and the breakdowns of the contributions of individual transitions, along with a comparison with some available theoretical results~\cite{dong2021,Themelis1995,Mitroy2010,Pouchan1998,Claude1998,Shukla2020,Patil2000,porsev2006,Komasa2002,Komasa2001,Bendazzoli2004, Sahoo2008,Singh2013,Mitroy2003,BANERJEE2010}. We can find that the polarizability of the $2s^2~^1S_0$ state is dominated by the $2s^2~^1S_0 \rightarrow 2s2p~^1P_1$ transition, while for the $2s2p~^3P_0$ state is dominated by the $2s2p~^3P_0 \rightarrow 2s3s~^3S_1$, $2s2p~^3P_0 \rightarrow 2p^2~^3P_1$, and $2s2p~^3P_0 \rightarrow 2s3d~^3D_1$ transitions. The ``Remains'' in the table represents the contributions from highly excited bound and continuum states of the valence electrons. The ``Core'' denotes the contributions of the core ($1s^2$) electrons, which is calculated by using a pseudospectral oscillator strength distribution ~\cite{Margoliash1978,Mitroy2003,Kumar1985}. The present total polarizability is in good agreement with other theoretical results~\cite{dong2021,Themelis1995,Mitroy2010, Pouchan1998,Claude1998, Patil2000,porsev2006,Komasa2002,Komasa2001,Bendazzoli2004,Sahoo2008, Singh2013,Mitroy2003,BANERJEE2010,Figari1983}, and the difference is no more than 1\%. 

%\subsection{Dynamic E1 polarizabilities}

\begin{figure*}[!h]
	\centering
	\includegraphics[width=10cm,height=8.5cm]{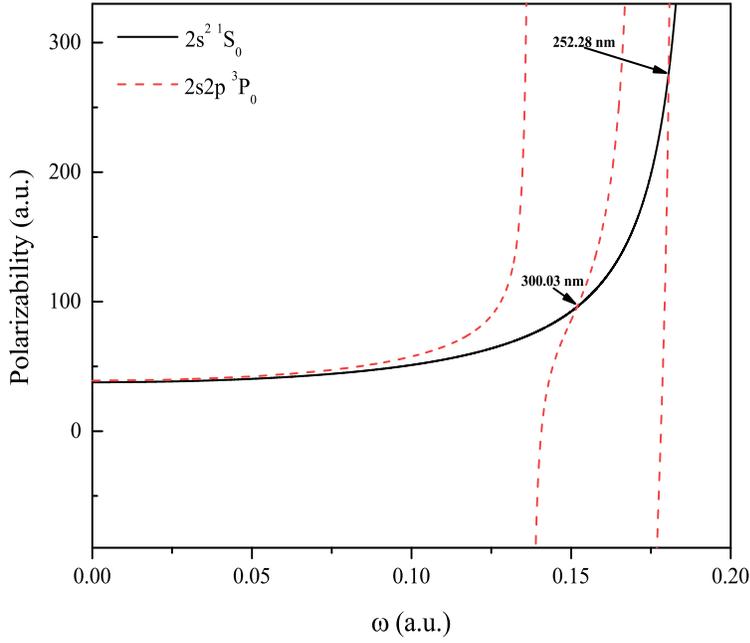}
	\captionsetup{font=footnotesize,labelfont=bf}
	\caption{\label{Fig1}Dynamic $E1$ polarizabilities (a.u.) of the $2s^2~^1S_0$ and $2s2p~^3P_0$ states of Be atoms. Two magic wavelengths are identified with arrows.}
\end{figure*}

Fig.~\ref{Fig1} depicts the dynamic polarizabilities of the $2s^2~^1S_0$ and $2s2p~^3P_0$ states. Two magic wavelengths are found which are identified with arrows. One of them, 300.03 nm, lies between the resonant transitions of $2s2p~^3P_0 \rightarrow 2s3s~^3S_1$ and $2s2p~^3P_0 \rightarrow 2p^2~^3P_1$. Another one, 252.28 nm, is located near the resonant wavelength of $2s2p~^3P_0 \rightarrow 2p^2~^3P_1$ transition. The present results are in good agreement with the calculations from Ref.~\cite{Mitroy2010}, 300.2 and 252.3 nm. Here, we recommend that the 300.03-nm magic wavelength can be used for magic-wavelength trapping in the experiment, since this magic wavelength has a 30-nm difference from the resonant wavelength of $2s2p~^3P_0 \rightarrow 2s3s~^3S_1$ transition. However, the 252.28-nm magic wavelength is only a 10-nm difference from the resonant wavelength of $2s2p~^3P_0 \rightarrow 2p^2~^3P_1$ transition, and it is far away from the visible region. Therefore, it is the best choice to use the 300.03-nm magic wavelength for magic-wavelength trapping in experiments.

Table~\ref{dynamicE1} lists the contributions of individual transitions to the dynamic $E1$ polarizabilities of the $2s^2~^1S_0$ and $2s2p~^3P_0$ states at the magic wavelength 300.03 nm. The polarizability of the $2s^2~^1S_0$ state is dominated by the resonant $2s^2~^1S_0 \rightarrow 2s2p~^1P_1$ transition, the contribution is more than 98\%, while the polarizability of the $2s2p~^3P_0$ state is dominated by the $2s2p~^3P_0 \rightarrow 2s3s~^3S_1$, $2s2p~^3P_0 \rightarrow 2p^2~^3P_1$, and $2s2p~^3P_0 \rightarrow 2s3d~^3D_1$ transitions. The contribution of $2s2p~^3P_0 \rightarrow 2s3s~^3S_1$ transition is negative. 

\begin{table}[ht]
	\centering
	\footnotesize\rm
	\captionsetup{font=footnotesize,labelfont=bf}
		\caption{\label{dynamicE1} Breakdowns of the contributions of individual
			transitions to the dynamic $E1$ polarizabilities (a.u.) of the $2s^2~^1S_0$ and
			$2s2p~^3P_0$ states at the 300.03 nm magic wavelength.}
	\setlength{\tabcolsep}{8.5mm}{
		\begin{tabular}{crccr}
			\\\hline  \hline
			\multicolumn{2}{c}{	$2s^2~^1S_0$}&   & \multicolumn{2}{c}{$2s2p~^3P_0$}   \\ \cline{1-2}  \cline{4-5}	Contributions&            $\alpha$&         &        Contributions&           $\alpha$       \\ \hline
			$2s2p~^1P_1$&       94.6108&         &  $2s3s~^3S_1$&          $-$19.4274       \\
			$2s3p~^1P_1$&        0.1846&         &  $2p^2~^3P_1$&             69.0677       \\ 
			&              &                     &  $2s3d~^3D_1$&             28.7577       \\
			&              &                     &  $2s4d~^3D_1$&              4.6003        \\
			Remainds&        1.1508&                  &          &             12.9479        \\  
			Core&        0.0523&                  &     &              0.0523        \\
			Total&       95.9985&                  &     &             95.9985        \\\hline  \hline
		\end{tabular}}
\end{table}

\subsection{Static $M1$ and $E2$ polarizabilities as well as hyperpolarizabilities}

The static $M1$ and $E2$ polarizabilities for the 
state $i$ can be given by~\cite{Porsev2004}
\begin{equation}
\label{M1}
\alpha^{M1}= \frac{2}{3(2j_{i}+1)}\sum\limits_{n}
\frac{|\langle \beta_{n}j_{n}\|\bm{M1}\
	|\beta_{i}j_{i}\rangle |^2}{\Delta E_{n \rightarrow i}},
\end{equation}
\begin{equation}
\label{E2}
\alpha^{E2}= \frac{2}{5(2j_{i}+1)}\sum\limits_{n}
\frac{|\langle \beta_{n}j_{n}\|\bm{Q}\
	|\beta_{i}j_{i}\rangle |^2}{\Delta E_{n \rightarrow i}},
\end{equation}
where $\bm{M1}$ and $\bm{Q}$ are the magnetic-dipole and
electric-quadrupole transition operators, respectively.

The dynamic hyperpolarizabilities $\gamma^l(\omega)$ and $\gamma^c(\omega)$
under the linearly and circularly polarized lights for the $j_{i}=0$
state can be written as, respectively~\cite{Porse2018,Tang2004}
\begin{equation}
\label{l}
\gamma^l(\omega)= \frac{1}{9}\mathcal{T}(1,0,1,\omega,-\omega,\omega)+\frac{2}{45}\mathcal{T}(1,2,1,\omega,-\omega,\omega),
\end{equation}
\begin{equation}
\label{c}
\gamma^c(\omega)= \frac{1}{9}\mathcal{T}(1,0,1,\omega,-\omega,\omega)+\frac{1}{90}\mathcal{T}(1,2,1,\omega,-\omega,\omega).
\end{equation}
Where $\mathcal{T}(1,0,1,\omega,-\omega,\omega)$ and $\mathcal{T}(1,2,1,\omega,-\omega,\omega)$ are 
expressed as the following general formula~\cite{Tang2004}
	\begin{align}
	&\mathcal{T}(j_{a},j_{b},j_{c},\omega,-\omega,\omega) = 4\sum\limits_{\beta_{a},\beta_{b},\beta_{c}}
	\langle \beta_{i} j_{i}||D||\beta_{a}j_{a}\rangle \langle \beta_{a}j_{a}||D||\beta_{b}j_{b}\rangle
	\langle \beta_{b}j_{b}||D||\beta_{c}j_{c}\rangle \langle \beta_{c}j_{c}||D||\beta_{i} j_{i} \rangle \nonumber \\
	&\times \{ \frac{1}{(\Delta E_{a \rightarrow i}-\omega)(\Delta E_{b \rightarrow i}-2\omega)(\Delta E_{c \rightarrow i}-\omega)}+\frac{1}{(\Delta E_{a \rightarrow i}+\omega)(\Delta E_{b \rightarrow i}+2\omega)(\Delta E_{c \rightarrow i}+\omega)} \nonumber \\
	&+ \frac{4\Delta E_{a \rightarrow i}\Delta E_{c \rightarrow i}}{(\Delta E_{a \rightarrow i}+\omega)(\Delta E_{a \rightarrow i}-\omega)\Delta E_{b \rightarrow i}(\Delta E_{c \rightarrow i}+\omega)(\Delta E_{c \rightarrow i}-\omega)} \}  + 8(-1)^{j_{a}+j_{c}+1}\delta(j_{b},j_{i})  \nonumber \\
	&\times \sum\limits_{\beta_{a}} \frac{\Delta E_{a \rightarrow i}\left| \langle \beta_{i} j_{i}||D||\beta_{a}j_{a}\rangle \right|^{2}}{\Delta E_{a \rightarrow i}^2-\omega^2} \sum\limits_{\beta_{c}} \frac{(3\Delta E_{c \rightarrow i}^2+\omega^2)\left| \langle \beta_{i} j_{i}||D||\beta_{c}j_{c}\rangle \right|^{2}}{(\Delta E_{c \rightarrow i}^2-\omega^2)^2},
	\end{align}
where $\sum\limits_{\beta_{a},\beta_{b},\beta_{c}}$ represents three summations over a large number of intermediate states. When $\omega$ = 0, Eqs. (\ref{l}) and (\ref{c}) are reduced to the static hyperpolarizabilities.
	
Table~\ref{M1E2} lists the presently calculated the static $M1$ and $E2$ polarizabilities for the $2s^2~^1S_0$ and $2s2p~^3P_0$ states. These polarizabilities are compared with some available theoretical results ~\cite{Komasa2001,Jiang2013J,porsev2006,Pouchan1998,Figari1983,Thakkar1989}. For the $E2$ polarizability of the $2s^2~^1S_0$ state, the present result is in excellent agreement with the calculation of the explicitly correlated Gaussian (ECG) basis~\cite{Komasa2001}, the CICP~\cite{Jiang2013J}, and the configuration interaction approach and many-body perturbation theory (CI+MBPT)~\cite{porsev2006}. The difference is no more than 0.15\%. There are no other theoretical or experimental values available for the $E2$ polarizability of the $2s2p~^3P_0$ state and the $M1$ polarizabilities of the $2s^2~^1S_0$ and $2s2p~^3P_0$ states.

\begin{table}[ht]
	\centering
	\footnotesize\rm
	\captionsetup{font=footnotesize,labelfont=bf}
	\caption{\label{M1E2} Static $M1$ and $E2$ polarizabilities (a.u.) for
		the $2s^2~^1S_0$ and $2s2p~^3P_0$ states.}
	\setlength{\tabcolsep}{5.5mm}{
		\begin{tabular}{cccccc}
			\hline  \hline
			\multicolumn{1}{c}{\multirow{2}*{Methods}}& \multicolumn{2}{c}{$2s^2~^1S_0$}& & \multicolumn{2}{c}{$2s2p~^3P_0$}   \\ \cline{2-3}  \cline{5-6}
			&              $\alpha^{E2}$&              $\alpha^{M1}$&         &        $\alpha^{E2}$&              $\alpha^{M1}$                      \\ \hline
			This work&                    300.98&          3.68$\times$ 10$^{-7}$&   &           1.58$\times$ 10$^6$&         4.87$\times$ 10$^5$\\
			ECG~\cite{Komasa2001}&                300.96&                            &    & &                 \\ 
			CICP~\cite{Jiang2013J}&                300.7&                            &   &   &                \\ 
			CI+MBPT~\cite{porsev2006}&           300.6(3)&                          &   &    &                  \\
			TDG1~\cite{Pouchan1998}&             285.6&                             &   &    &              \\ 
			VP+CI~\cite{Figari1983}&               299.4&                             &    &  &               \\
			CCD+ST~\cite{Thakkar1989}&              298.8&                             &   &  &    \\\hline  \hline
		\end{tabular} }
	\end{table}

Table~\ref{hy} presents the static hyperpolarizabilities of the $2s^2~^1S_0$ and $2s2p~^3P_0$ states. We find that the static hyperpolarizabilities are dominated by the $\mathcal{T}(1,2,1)$ term. For the $2s^2~^1S_0$ state, the present RCICP result is in good agreement with other theoretical results~\cite{Koch1991,Thakkar1989,Tunega1997,Strasburger2014,Papadopoulos1995,Stiehler1995,Pluta1992,Kobus2015}. There are no other theoretical values for the $2s2p~^3P_0$ state available for comparison.

\begin{table}[ht]
	\centering
	\footnotesize\rm
	\captionsetup{font=footnotesize,labelfont=bf}
	\caption{\label{hy} The static hyperpolarizabilities $\gamma$ (a.u.)
		of the $2s^2~^1S_0$ and $2s2p~^3P_0$ states.
		The notation a[b] means a$\times10^{b}$.}
	\setlength{\tabcolsep}{10.5mm}{
		\begin{tabular}{cccc}
				\hline  \hline
	Methods	&	                     Contributions&      $2s^2~^1S_0$&       $2s2p~^3P_0$  \\ \hline
	        &   $\frac{1}{9}\mathcal{T}(1,0,1)$&     6.081[3] &        4.904[4]  \\
			&   $\frac{2}{45}\mathcal{T}(1,2,1)$&    2.963[4]&        2.628[5]  \\
   This work&                                   &    3.571[4]&         3.118[5]    \\
FCI~\cite{Koch1991}&                            &    2.7227[4]&                      \\
CCD+ST~\cite{Thakkar1989}&                      &    3.148[4& \\
MP4~\cite{Papadopoulos1995}&                    &    3.1[4]& \\
CC-R12~\cite{Tunega1997}&                       &    3.21[4]& \\
ECG~\cite{Strasburger2014}&                     &    3.0989[4]& \\
RHF~\cite{Stiehler1995}&                        &    3.95[4]& \\
MCSCF~\cite{Pluta1992}&                         &    3.930[4]& \\
FD HF~\cite{Kobus2015}&                         &    3.912[4]   	\\\hline  \hline
		\end{tabular}  }                                              
\end{table}	

\subsection{Dynamic $E2$ and $M1$ polarizabilities as well as dynamic hyperpolarizabilities around magic wavelength.}

The dynamic $M1$ and $E2$ polarizabilities for the $j_{i}=0$
state can be given by~\cite{Porsev2004}
\begin{equation}
\label{dM1}
\alpha^{M1}(\omega)= \frac{2}{3}\sum\limits_{n}
\frac{\Delta E_{n \rightarrow i}|\langle \beta_{n}j_{n}\|\bm{M1}\
	|\beta_{i}j_{i}\rangle |^2}{\Delta E^2_{n \rightarrow i}-\omega^2},
\end{equation}
and
\begin{equation}
\label{dE2}
\alpha^{E2}(\omega)= \frac{1}{30}(\alpha\omega)^2\sum
\limits_{n}\frac{\Delta E_{n \rightarrow i}|\langle
	\beta_{n}j_{n}\|\bm{Q}\|\beta_{i}j_{i}\rangle |^2}
{\Delta E^2_{n \rightarrow i}-\omega^2},
\end{equation}
where $\alpha$ in Eq.~(\ref{dE2}) is the fine structure constant.

Table~\ref{dynamicM1E2} lists the presently calculated dynamic $M1$ and $E2$ polarizabilities for the $2s^2~^1S_0$ and $2s2p~^3P_0$ states at the 300.03-nm magic wavelength. As can be seen from the table, the absolute value of $M1$ polarizability of the $2s^2~^1S_0$ state is five orders of magnitude smaller than $2s2p~^3P_0$ state, and the $M1$ polarizability of the $2s2p~^3P_0$ state is negative. Thus, the differential $M1$ polarizability ($\Delta\alpha^{M1}(\omega)$) between these two states is determined by the $2s2p~^3P_0$ state. The differential $E2$ polarizability ($\Delta\alpha^{E2}(\omega)$) is one order of magnitude larger than that of the $\Delta\alpha^{M1}(\omega)$. Therefore, the differential dynamic multipolar polarizability ($\Delta\alpha^{QM}(\omega)=\Delta\alpha^{M1}(\omega)+\Delta\alpha^{E2}(\omega)$) is mainly determined by the $\Delta\alpha^{E2}(\omega)$. 

\begin{table}[ht]
	\centering
	\footnotesize\rm
	\captionsetup{font=footnotesize,labelfont=bf}
	\caption{\label{dynamicM1E2} The dynamic $M1$ and $E2$ polarizabilities (a.u.) 
		for the $2s^2~^1S_0$ and $2s2p~^3P_0$ states at the 300.03-nm
		magic wavelength. The dynamic multipolar polarizability $\alpha^{QM}(\omega)=\alpha^{M1}(\omega)+
		\alpha^{E2}(\omega)$. The $\Delta\alpha$ represents the differential polarizabilities
		between these two states. The notation a[b] means a$\times10^{b}$.}
	\setlength{\tabcolsep}{8.5mm}{
		\begin{tabular}{cccc}
				\hline  \hline
			Polarizabilities&            $2s^2~^1S_0$&             $2s2p~^3P_0$&             $\Delta\alpha(2s^2~^1S_0\rightarrow 2s2p~^3P_0)$              \\ \hline
			$\alpha^{M1}(\omega)$&          2.38[$-$9]&                $-$1.47[$-$4]&                  $-$1.47[$-$4]                                          \\
			$\alpha^{E2}(\omega)$&          4.33[$-$4]&                 2.85[$-$3]&                       2.42[$-$3]                                          \\
			$\alpha^{QM}(\omega)$&          4.33[$-$4]&                 2.70[$-$3]&                       2.27[$-$3]                          \\	\hline  \hline		
		\end{tabular}  }                                             
\end{table}

Table~\ref{hyper} gives the dynamic hyperpolarizabilities of the $2s^2~^1S_0$ and $2s2p~^3P_0$ states and breakdowns of the contributions to the dynamic hyperpolarizabilities at 300.03 nm magic wavelength. We found that the dynamic hyperpolarizability $\gamma^l(\omega)$ of the $2s^2~^1S_0$ state in the linearly polarized light is four orders of magnitude smaller than $2s2p~^3P_0$ state, and the $\gamma^c(\omega)$ of the $2s^2~^1S_0$ state in the circularly polarized light is three orders of magnitude  smaller than the $2s^2~^3P_0$ state. Therefore, the differential dynamic hyperpolarizabilities ($\Delta\gamma^l(\omega)$ and $\Delta\gamma^c(\omega)$) in the linearly and circularly polarized lights are determined by the $2s2p~^3P_0$ state.

\begin{table}[ht]
	\centering
	\footnotesize\rm
	\captionsetup{font=footnotesize,labelfont=bf}
	\caption{\label{hyper} The dynamic hyperpolarizabilities (a.u.)
		of the $2s^2~^1S_0$ and $2s2p~^3P_0$ states at the 300.03-nm magic wavelength.
		The $\Delta\gamma{(\omega)} $ represents the differential hyperpolarizabilities
		between these two states. The superscript $l$ and $c$ represent the linearly
		and circularly polarized lights. The notation a[b] means a$\times10^{b}$.}
	\setlength{\tabcolsep}{3.8mm}{
		\begin{tabular}{cccccc}
				\hline  \hline
			\multicolumn{1}{c}{\multirow{2}*{Contribution}}&	\multicolumn{2}{c}{$\gamma^l{(\omega)}$}& \multicolumn{1}{c}{\multirow{2}*{Contribution}}& \multicolumn{2}{c}{$\gamma^c{(\omega)}$}   \\ \cline{2-3}  \cline{5-6}
			&                                       $2s^2~^1S_0$&       $2s2p~^3P_0$&                                                   &     $2s^2~^1S_0$&       $2s2p~^3P_0$           \\ \hline    
			$\frac{1}{9}T(1,0,1,\omega,$-$\omega,\omega)$&           $-$7.91[5]&      $-$3.01[6]&           $\frac{1}{9}T(1,0,1,\omega,$-$\omega,\omega)$&       $-$7.91[5]&      $-$3.01[6]  \\
			$\frac{2}{45}T(1,2,1,\omega,$-$\omega,\omega)$&              7.65[5]&        7.43[8]&           $\frac{1}{90}T(1,2,1,\omega,$-$\omega,\omega)$&          1.91[5]&        1.86[8]  \\
			Total&                                     $-$2.59[4]&        7.40[8]&    &       $-$6.00[5]&        1.83[8]  \\
			$\Delta\gamma^l{(\omega)}$&                                    7.40[8]&              &                              $\Delta\gamma^c{(\omega)}$&          1.83[8]    \\ 	\hline  \hline 
		\end{tabular}    }                                          
\end{table}
\begin{figure*}[!h]
	\centering
	\includegraphics[width=14cm,height=10cm]{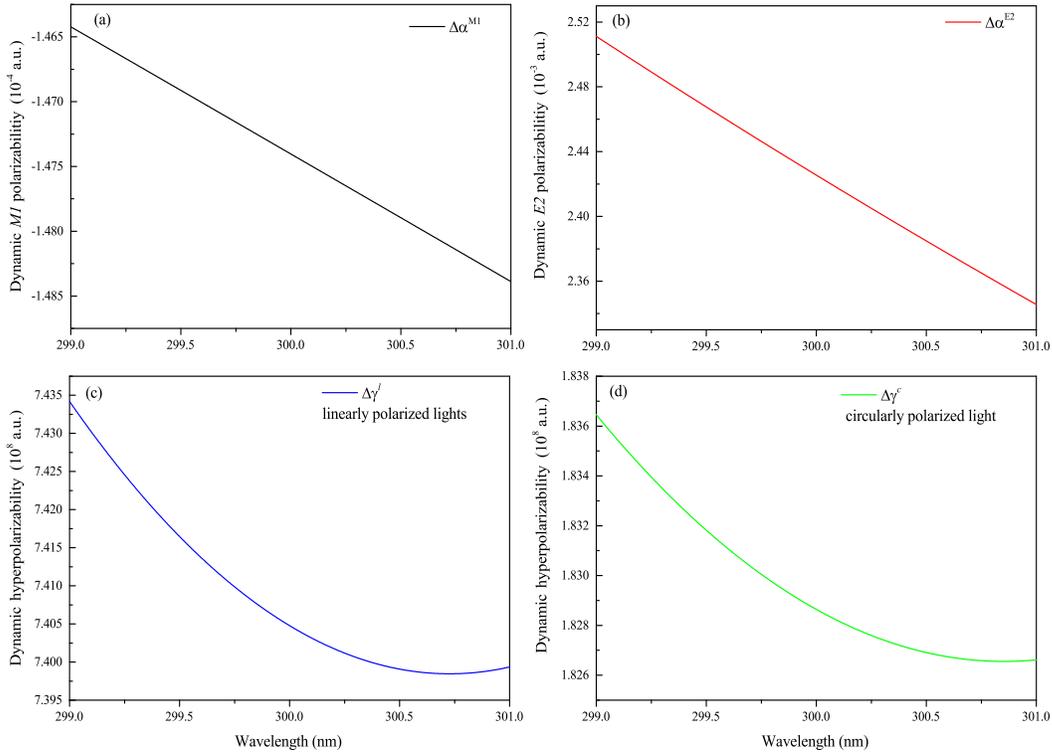}
	\captionsetup{font=footnotesize,labelfont=bf}
	\caption{\label{Fig2} Differential dynamic $M1$ and $E2$ polarizabilities (a.u.) as well as hyperpolarizabilitiy (a.u.) around the 300.03-nm magic wavelength. (a-b) The differential dynamic $M1$ and $E2$ polarizabilities between $2s^2~^1S_0$ and $2s2p~^3P_0$ states. (c-d) The differential dynamic hyperpolarizabilities between $2s^2~^1S_0$ and $2s2p~^3P_0$ states in the linearly and circularly polarized lights.}
\end{figure*}

Fig.~\ref{Fig2} shows the differential dynamic $M1$ and $E2$ polarizabilities as well as hyperpolarizabilities between $2s^2~^1S_0$ and $2s2p~^3P_0$ states around the 300.03-nm magic wavelength. These differential polarizabilities are extremely important in analyzing multipole and nonlinear Stark shifts.

\subsection{Stark shifts near the operational magic conditions}

For atoms trapped under a one-dimensional optical lattice
with the laser frequency $\omega$ and the linearly polarized laser field
intensity $I$, the Stark shift for a clock transition can be expressed as~\cite{Katori2015,Ushijima2018}
\begin{align}
\label{stark}
& h\Delta\nu=[\frac{\partial\Delta\alpha^{E1}(\omega)}{\partial\nu}\delta-\Delta\alpha^{QM}(\omega)](n_{z}+\frac{1}{2})
\sqrt{\frac{E_{R}}{\alpha^{E1}(\omega)}}I^{1/2} \nonumber \\ &-[\frac{\partial\Delta\alpha^{E1}(\omega)}{\partial\nu}+
\frac{3}{8}\frac{E_{R}\Delta\gamma^{l}(\omega)}{\alpha^{E1}(\omega)}(n_{z}^2+n_{z}+\frac{1}{2})]I \nonumber \\ &
+\frac{1}{2}\Delta\gamma^{l}(\omega)\sqrt{\frac{E_{R}}{\alpha^{E1}(\omega)}}(n_{z}+\frac{1}{2})I^{3/2}-\frac{1}{4}\Delta\gamma^{l}(\omega)I^{2},
\end{align}
where $\delta$ is the frequency detuning of the lattice laser frequency $\nu$ relative to the magic frequency $\nu_{m}=\omega_{m}/2\pi$, $n_{z}$ is the vibrational state of atoms along the $z$ axis~\cite{Taichenachev2008}, and $E_{R}=h^2/(2M\lambda_{magic}^2)$ is the lattice photon recoil energy with $M$ being the atomic mass. 

\begin{figure*}[!h]
	\centering
	\includegraphics[width=14cm,height=10cm]{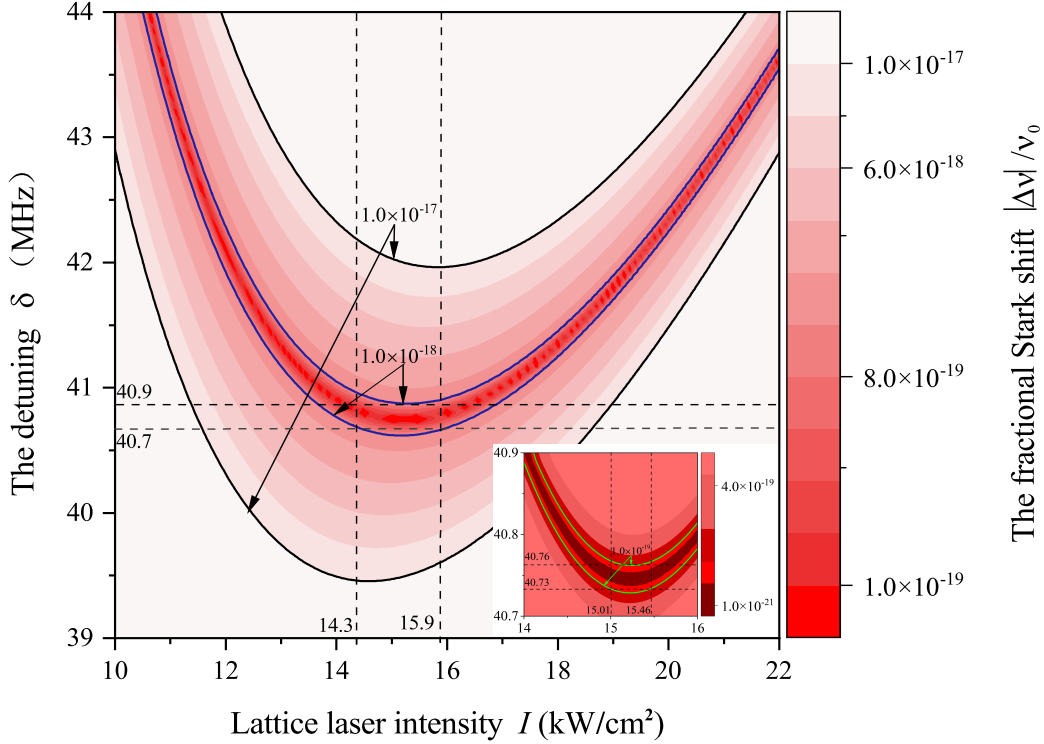}
	\captionsetup{font=footnotesize,labelfont=bf}
	\caption{\label{Fig3}The fractional Stark shifts $\left| \Delta\nu\right| /\nu_{0}$ of the clock transition with the increase of laser intensity $I$ and detunings $\delta$. The color gradients represent the different fractional Stark shifts. The black solid lines indicate the contour line of the fractional Stark shifts of 1.0 $\times$ 10$^{-17}$. The blue solid lines indicate the contour line of fractional Stark shifts of 1.0 $\times$ 10$^{-18}$, and when the laser intensity $I$ is in the range of 14.3 $\sim$ 15.9 kW/cm$^{2}$ and detuning $\delta$ is in the range of 40.7 $\sim$ 40.9 MHz, the fractional Stark shift is less than the level of 1.0 $\times$ 10$^{-18}$. The green solid lines in the illustration represent the contour line of the fractional Stark shifts of 1.0 $\times$ 10$^{-19}$, and when the $I$ is in the range of 15.01 $\sim$ 15.46 kW/cm$^{2}$ and $\delta$ is in the range of 40.73 $\sim$ 40.76 MHz, the fractional Stark shift is less than the level of 1.0 $\times$ 10$^{-19}$.}
\end{figure*}

Here, we assume that the Be atoms are trapped in the $n_{z} = 0$ vibrational state. The multipolar and nonlinear Stark shifts are obtained using Eq.~(\ref{stark}). 
Fig.~\ref{Fig3} presents the fractional Stark shifts $\left| \Delta\nu\right| /\nu_{0}$ of clock transition, the ratio of absolute values of Stark shifts to clock transition frequency, with the increase of laser intensity $I$ and detuning $\delta$. The color gradients represent the different fractional Stark shifts. The black solid lines indicate the contour line of fractional Stark shifts of 1.0 $\times$ 10$^{-17}$, and the blue solid lines indicate the contour line of fractional Stark shifts of 1.0 $\times$ 10$^{-18}$. The green solid lines in the illustration represent the contour line of fractional Stark shifts of 1.0 $\times$ 10$^{-19}$.
In order to reduce the multipolar and nonlinear Stark shifts and make the Stark shifts insensitive to the $I$ and $\delta$, it should choose the vertex of the contour lines to determine the position of the laser intensity $I$ and detuning $\delta$. We find that when the $I$ is in the range of 14.3 $\sim$ 15.9 kW/cm$^{2}$ and $\delta$ is in the range of 40.7 $\sim$ 40.9 MHz, the fractional Stark shifts of the clock transition are lower than 1.0 $\times$ 10$^{-18}$. While, when the $I$ is in the range of 15.01 $\sim$ 15.46 kW/cm$^{2}$ and $\delta$ is in the range of 40.73 $\sim$ 40.76 MHz, the fractional Stark shifts are lower than 1.0 $\times$ 10$^{-19}$, as shown in the illustration in Fig.~\ref{Fig3}. These distinctive conditions can provide a reference for the development of the Be optical lattice clock at the level of 10$^{-19}$.

\section{Conclusions}

The energy levels and $E1$ matrix elements of the low-lying states of Be atoms have been calculated using the RCICP method. The static and dynamic $E1$, $M1$, and $E2$ polarizabilities as well as hyperpolarizabilities of the $2s^2~^1S_0$ and $2s2p~^3P_0$ states are determined. Then, two magic wavelengths, 300.03 nm and 252.28 nm, of the $2s^2~^1S_0 \rightarrow 2s2p~^3P_0$ clock transition are found. We recommend that the 300.03-nm magic wavelength can be used for magic trapping.  The $\Delta\alpha^{M1}(\omega)$ and $\Delta\alpha^{E2}(\omega)$ as well as differential dynamic hyperpolarizabilities around the 300.03-nm magic wavelength are determined. 
In addition, we find that the $\Delta\alpha^{M1}(\omega)$ is determined by $2s2p~^3P_0$ state. The $\Delta\alpha^{QM}(\omega)$ is mainly determined by the $\Delta\alpha^{E2}(\omega)$. The differential dynamic hyperpolarizability in the linearly and circularly polarized lights are all determined by the $2s2p~^3P_0$ state. 

Finally, the multipolar and nonlinear Stark shifts of the clock transition near the magic wavelength are calculated in detail. We find that when the laser intensity is in the range of 14.3 $\sim$ 15.9 kW/cm$^{2}$ and $\delta$ is in the range of 40.7 $\sim$ 40.9 MHz, the fractional Stark shifts of the clock transition are lower than 1.0 $\times$ 10$^{-18}$. While, when the $I$ is in the range of 15.01 $\sim$ 15.46 kW/cm$^{2}$ and $\delta$ is in the range of 40.73 $\sim$ 40.76 MHz, the fractional Stark shifts are lower than 1.0 $\times$ 10$^{-19}$. These will provide important support for developing ultra-high-precision Be optical clocks.

\section*{Acknowledgments}
This work has been supported by the National Natural Science Foundation of China under Grants No. 12174316, the Young Teachers Scientific Research Ability Promotion Plan of Northwest Normal University (NWNU-LKQN2020-10) and Funds for Innovative Fundamental Research Group Project of Gansu Province (20JR5RA541).

%\begin{thebibliography}{10}
%\bibitem{book1} Goosens M, Rahtz S and Mittelbach F 1997 {\it The \LaTeX\ Graphics Companion\/}
%(Reading, MA: Addison-Wesley)
%\bibitem{eps} Reckdahl K 1997 {\it Using Imported Graphics in \LaTeX\ } (search CTAN for the file `epslatex.pdf')
%\end{thebibliography}
%%\bibliographystyle{myiopart-num}
%\bibliographystyle{iopart-num}


%\bibliographystyle{myiopart-num}

%\bibliography{wu}
\providecommand{\newblock}{99}

\end{document}